\newcommand{\bld}[1]{\mbox{\boldmath{$#1$}}}

\documentstyle[preprint,aps,12pt,floats]{revtex}
        
\begin{document}
\draft
\tighten

\setcounter{secnumdepth}{2}

\renewcommand{\floatpagefraction}{0.7}

\begin{titlepage}
\hfill \parbox{5cm}{\raggedleft TTP96--16\\atom-ph/9605005\\ Mai 1996}

\vspace*{1.5cm}
\begin{center}{\Large HYPERFINE INTERACTIONS\\
\vspace{0.3cm} BETWEEN ELECTRONS}\\
\vspace{1cm}
\large{Ruth H\"ackl and Hartmut Pilkuhn}\\
\vspace{0.5cm}
{\em Institut f\"ur Theoretische Teilchenphysik,\\
Universit\"at Karlsruhe,\\ Postfach 6980,\\
D--76228 Karlsruhe, Germany}
\end{center}
\vspace{8cm}
--------\\
{\footnotesize \begin{tabular}{ll}
Internet : & Ruth.Haeckl@physik.uni-karlsruhe.de\\
 & hp@ttpux2.physik.uni-karlsruhe.de \end{tabular}}
\end{titlepage}
\begin{abstract}{
The relativistic Breit Hamiltonian between electrons is transformed into
an effective vector potential ${\bf A}_i$ for the $i.$th
electron, ${\bf A}_i$ having the structure of a recoil--corrected
hyperfine operator.
 Apart from a small three--body operator, the Dirac--Breit equation
is now easier applied to relativistic magnetic properties of complex systems.
\vspace{1.8cm}
}
\end{abstract}

Relativistic $n$--electron systems of atoms or molecules are quite
precisely described by the Dirac--Breit equation \cite{grant}.
In addition to the ordinary Dirac--Coulomb Hamiltonian, the equation
contains the Breit Hamiltonian \cite{breit}
\begin{equation}\label{gl1}
 H_{B}=\sum_{i<j}  B_{ij}\,  , \qquad  B_{ij}=
-e^2 (\bld{\alpha}_i \bld{\alpha}_j + \alpha _{ir} 
\alpha _{jr})/2 r_{ij}\, ,
\end{equation}
where the $\bld{\alpha}$ are Dirac matrices and $\alpha_{ir}$ and 
$\alpha_{jr}$ are the components of $\bld{\alpha}_{i}$ and 
$\bld{\alpha}_{j}$
along the direction $\hat{\bf r}$ of ${\bf r}_i -{\bf r}_j$.
$H_{B}$ contributes to the electronic fine structure. The term
hyperfine interaction is normally reserved for the interaction with
the nuclear spin \cite{bethe}.

Breit operators are somewhat inconvenient. Firstly, ${H}_{B}$ is the
only operator in the Dirac--Breit equation which exchanges large and small 
components of two electrons simultaneously. Secondly, the use of 
${H}_{B}$ beyond first--order perturbation theory requires complicated
{\em improved versions} of ${B}_{ij}$ \cite{lindroth}. This is so 
because ${B}_{ij}^2$ is too large. 

In this letter we present a transformation which removes both inconveniences.
One of the new operators produced by the transformation is complicated but
should be negligible. Without it, the Dirac--Breit equation for
$n$ electrons of total energy $E$ with a total potential
$V=\sum V({\bf r}_i) + \sum_{i<j} e^2 /r_{ij}$
becomes
\begin{equation}\label{gl2}
\left\lbrack E -V - mc^2 \sum \beta_{i} - \sum \bld{\alpha}_{i} 
( c {\bf p}_{i} + e {\bf A}_{0i} + e {\bf A}_{{\rm{\tiny hf}},i} ) 
\right\rbrack 
\Psi =0\, .
\end{equation}
Here ${\bf A}_{0i}$ is the vector potential produced by spinless
electrons,
\begin{equation}\label{gl3}
{\bf A}_{0i} =  - \frac{e}{4mc} \sum_{j\ne i}  r_{ij} ^{-1} ({\bf p}_{j}
+ \hat{{\bf r}}_{ij} p_{jr} )
\end{equation}
and ${\bf A}_{{\rm{\tiny hf}},i}$ is the vector potential of the electron spins
\begin{equation}\label{gl4}
{\bf A}_{{\rm{\tiny hf}},i} =- \frac{e }{4mc} 
\nabla _i \times \sum_{j\ne i}
\bld{\sigma}_j / r_{ij}\, .
\end{equation}
This is nothing but the ordinary hyperfine interaction with spinor
particles $j$ of $g$--factors $g_j =2$ and masses $m_j$, for the
special case $m_i=m_j=m$. If one is only interested in spin--polarized 
systems, ${\bf A}_{0i}$ may also be neglected.

The operator ${\bf p }_i = - i \hbar \nabla_{i}$ is the canonical
momentum. The momentum which enters the kinetic energy is $\bld{\pi}_{i} =
{\bf p}_{i} +e{\bf A}_{i}/c\,$. One may therefore say that the Breit
operator has become part of the kinetic energy. This form could be useful
e.g.~for relativistic density functional theories, in which  the ${\bf A}$
appearing in the Kohn--Sham equations represents more than an 
external magnetic field \cite{rajagopal}.

For the derivation of (\ref{gl2}) from the Dirac--Breit equation it is 
convenient to decompose $\bld{\alpha}=\gamma_{5}\bld{\sigma}$, where
the $\bld{\sigma}$ are Pauli matrices; $\gamma_5$ exchanges large and small
components. The $B_{ij}$ of (\ref{gl1}) becomes
\begin{equation}\label{gl5}
B_{ij} = -\gamma_{5i}\, \gamma_{5j}\,b_{ij}\ , \qquad b_{ij}=
(\bld{\sigma}_{i}\bld{\sigma}_{j} + \sigma_{ir} \sigma_{jr} )
 e^2/2 r_{ij}\, .
\end{equation}
As $\gamma_{5}^2 =1$, one finds $B_{ij}^2=b_{ij}^2$, which
is unacceptably large \cite{bethe}. The improved versions of $B_{ij}$
\cite{lindroth} give much smaller $B_{ij}^2$. In the following
we simply keep $B_{ij}$ (\ref{gl5}) but put $B_{ij}^2 =0$ by hand
\cite{malvetti1}. We also abbreviate
\begin{equation}\label{gl6}
E-V=c \pi^{0}, \qquad \sum \bld{\alpha}_{i} {\bf p}_{i} = P
\end{equation}
which
gives the original Dirac--Breit equation the compact form
\begin{equation}\label{gl7}
(\pi^{0} - mc \sum \beta_{i}-P -H_{B}/c) \Psi_{DB} =0 \, .
\end{equation}
The main step in the derivation of (\ref{gl2}) from (\ref{gl7}) has been
done in 1990 for the special case of two--electron atoms \cite{malvetti2}.
One defines an operator $Z$ which satisfies
\begin{equation}\label{gl8}
\lbrace \pi^{0} - mc \sum \beta_{i}\, , Z \rbrace = H_{\rm B} / c \, ,
\end{equation}
sets $\Psi_{DB} =(1+Z)\Psi$, and  multiplies (\ref{gl7}) by $(1+Z)$. As the
solution $Z$ of (\ref{gl8}) is proportional to $B_{ij}$, one
neglects terms of order $Z^2$ and $Z\,B_{ij}$, thus obtaining
\begin{equation}\label{gl9}
(\pi^{0} - mc \sum \beta_{i} - P - \lbrace P , Z\rbrace \, ) \Psi = 0 \, .
\end{equation}
Remembering $\lbrace \beta ,\gamma_{5} \rbrace =0$, one finds the following
solution of (\ref{gl8}):
\begin{equation}\label{gl10}
Z=\sum_{i<j} {\rm B}_{ij}/2(c\pi^{0} - mc^2 \sum_{k}\!'' \, \beta_{k} )\, .
\end{equation}
The $''$ on the last sum means omission of terms with $k=i$ or $j$. For
systems with only two electrons, the denominator of $Z$ is $2c\pi^{0}$.
For $n>2$ electrons, $Z$ is complicated, but when $\lbrace P , Z \rbrace$
in (\ref{gl9}) is small, we may approximate $\beta_{k}=1$ and also $\pi^{0}
=n mc$, such that the whole bracket is simply $2mc^2$.

As $Z$ contains two factors $\gamma_{5}$ and $P$ contains one, $\lbrace P ,Z
\rbrace $ contains a triple sum with three such factors. When 
the $\gamma_{5}$ from $P$ is $\gamma_{5i}$ or $\gamma_{5j}$,
the relation $\gamma_{5}^2=1$ reduces the three factors 
$\gamma_{5}$ to one.
The remaining terms will be called $\lbrace P, Z \rbrace_{3}$:
\begin{equation}\label{gl11}
\lbrace P,Z\rbrace = - \sum_{i\ne j}\ \left\lbrace {\bf p}_{j}
\bld{\sigma}_{j}\, ,
b_{ij} \right\rbrace \gamma_{5i}/4mc^2 + \lbrace P,Z\rbrace_{3} \, ,
\end{equation}
\begin{equation}\label{gl12}
\lbrace P,Z\rbrace_{3} = - \sum_{i\ne j\ne k} b_{ij}\, {\bf p}_k\,
 \bld{\sigma}_{k}\,
\gamma_{5i}\, \gamma_{5j}\, \gamma_{5k}/2mc^2 .
\end{equation}
$\lbrace P , Z \rbrace _{3}$ is now a three--body operator which is small on 
account of its three factors $\gamma_{5}$ which exchange large and small
components. Hopefully, it may be neglected in applications.

It remains to evaluate the anticommutators $\lbrace 
{\bf p}_{j}\bld{\sigma}_{j},
b_{ij} \rbrace$ in (\ref{gl11}):
\begin{equation}\label{gl13}
\lbrace {\bf p}_{j} \bld{\sigma}_{j}, b_{ij} \rbrace = e^2 r_{ij}^{-1}
\lbrack \bld{\sigma}_{i} {\bf p}_{j} + \sigma_{ir}\, p_{jr} + 
(\bld{\sigma}_{i}
\times \bld{\sigma}_{j})_{r}\, /r_{ij} \rbrack  \, .
\end{equation}
The last operator in the square bracket produces 
${\bf A}_{{\rm {\tiny hf}},i}$ by means of
the identity \mathchardef\sigma="711B
$(\bld{\sigma}_{i}\times \bld{\sigma}_{j})_{r}/r^2 = 
\bld{\sigma}_{i} ( \nabla \times \bld{\sigma}_{j}/r )\, .$
This completes our derivation of equation (\ref{gl2}). It is worth
mentioning that the combination $\bld{\alpha}_{i}\, e {\bf A}_{0i}$ is the
Breit interaction of the $i.$th electron 
with spinless particles.
It appears  also for nuclei, spinless or not, with $2m$ replaced by 
$m_{i} + m_{j}$ in (\ref{gl3}), and with $-e$ replaced by $Z_{j}e\,$.
For a single nucleus $j=N$, one may use ${\bf p}_{N}=-\sum_{i'}{\bf p}_{i'}$
to remove the ${\bf p}_{i}$--contribution in ${\bf A}_{0i}$ by a small
shift of the distance $r_{iN}$, $r_{iN}=r_{i} + Z e^2/2(m+m_{N})c^2$ 
\cite{malvetti1}.

\section*{Acknowledgment}
This work has been supported by the {\it Deutsche Forschungsgemeinschaft.}


\begin{thebibliography}{99}

\bibitem{grant} Grant I P and Quiney H M 1988 {\it Adv. Atom. Mol. Phys.}
 {\bf 23} 37

\bibitem{breit} Breit G 1929 {\it Phys. Rev.} {\bf 34} 553; 1930 
 {\it Phys. Rev} {\bf 36} 383

\bibitem{bethe}Bethe H A Salpeter E E 1957 {\it Quantum Mechanics of One--
 and Two--Electron Atoms} (Berlin: Springer)

\bibitem{lindroth}Lindroth E and M{\aa}rtensson--Pendrill A--M 1989 {\it
 Phys. Rev.} {\bf 39 A} 3794


\bibitem{rajagopal}Rajagopal A K 1980 {\it Adv. Chem. Phys} {\bf 41} 59;
 Vignale G and Rasolt M 1992 {\it Phys. Rev.} {\bf 46B} 10232


\bibitem{malvetti1}Malvetti M and Pilkuhn H 1994 {\it Phys. Rep.} C {\bf 248} 1



\bibitem{malvetti2}Malvetti M and Pilkuhn H 1990 {\it J. Phys. B} {\bf 23}
 3469; Malvetti M 1990 thesis (Karlsruhe)

 
\end{thebibliography}
\end{document}